\documentclass[aps,prb,superscriptaddress,twocolumn,showpacs]{revtex4-1}
\usepackage{amsmath,amssymb,graphicx,braket,hyperref}
\usepackage[usenames,dvipsnames]{xcolor}

\newcommand{\be}{\begin{equation}}
\newcommand{\ee}{\end{equation}}
\newcommand{\bea}{\begin{eqnarray}}
\newcommand{\eea}{\end{eqnarray}}
\newcommand{\ba}{\begin{eqnarray*}}
\newcommand{\ea}{\end{eqnarray*}}

\newcommand{\beq}{\begin{equation}}
\newcommand{\eeq}{\end{equation}}

\newcommand{\bma}{\begin{pmatrix}}
\newcommand{\ema}{\end{pmatrix}}

\newcommand{\im}{\text{Im}}

\begin{document}

\title{Emergent Finite Frequency Criticality of Driven-Dissipative Correlated Lattice Bosons}
\author{Orazio Scarlatella}
\affiliation{Institut de Physique Th\'{e}orique, Universit\'{e} Paris Saclay, CNRS, CEA, F-91191 Gif-sur-Yvette, France}
\author{Rosario Fazio}
\affiliation{ICTP, Strada Costiera 11, 34151 Trieste, Italy}
\affiliation{NEST, Scuola Normale Superiore $\&$ Istituto Nanoscienze-CNR, I-56126, Pisa, Italy}
\author{Marco Schir\'o}
\affiliation{Institut de Physique Th\'{e}orique, Universit\'{e} Paris Saclay, CNRS, CEA, F-91191 Gif-sur-Yvette, France}

\begin{abstract} 
Critical points and phase transitions are characterized by diverging susceptibilities, reflecting the tendency of the system toward spontaneous symmetry breaking. Equilibrium statistical mechanics bounds these instabilities to occur at zero frequency, giving rise to static order parameters. In this work we argue that a prototype model of correlated driven-dissipative lattice bosons, of direct relevance for upcoming generation of circuit QED arrays experiments, exhibits a susceptibility sharply diverging at a finite non-zero frequency, which is an emerging scale set by interactions and non-equilibrium effects. In the broken-symmetry phase the corresponding macroscopic order parameter becomes non-stationary and oscillates in time without damping, thus breaking continuous time-translational symmetry. Our work, connecting breaking of time translational invariance to divergent finite frequency susceptibilities, which are of direct physical relevance, could potentially be extended to study other time-domain instabilities in non-equilibrium quantum systems, including Floquet time crystals and quantum synchronization.
\end{abstract}
\maketitle

\section{Introduction}
Classical and quantum phase transitions in systems at thermal equilibrium are characterized, according to the Landau paradigm, by the emergence of a static order parameter which spontaneously breaks a symmetry of the system, such as spin rotational invariance for magnetism or spatial translational invariance for crystals~\cite{Landau5,Sachdev}. The resulting criticality is described in terms of an instability of the normal symmetric phase, characterized by a singularity of a static susceptibility. For classical systems far away from thermal equilibrium, such as in presence of external forcing and dissipation, the variety of instabilities can be far richer, with both finite momentum and finite frequency modes going unstable and resulting in the formation of patterns, propagating fronts, spatio-temporal chaos, synchronization or other oscillatory behaviors~\cite{CrossHohenbergRMP93,CrossGreenside_Pattern,VanSaarloosPhyRep03}.
In the quantum world, the question of whether finite frequency modes can become critical, giving rise to time-domain instabilities of the quantum dynamics and to an associated breaking of time-translational invariance, is much less explored. Experimental breakthroughs have brought forth a number of platforms which naturally probe non-equilibrium quantum many body dynamics, ranging from ultra cold atoms~\cite{BlochDalibardNascimbeneNatPhys12}, trapped ions~\cite{BlattRoosNatPhys12} and arrays of non-linear circuit QED cavities~\cite{Wallraf_Nature2004,AndrewNatPhys,SchmidtKochAnnPhy13,LeHurReview16}, thus making the question of experimental relevance. 
In this respect, quantum many body systems in presence of both driving and dissipation mechanisms~\cite{SiebereHuberAltmanDiehlPRL13,MaghrebiGorshkovPRB16,MarinoDiehlPRL16} represent natural platforms to understand and explore such time-domain dynamical instabilities. 
A well know example is provided by exciton-polariton condensates  where superfluidity emerges with an order parameter oscillating in time~\cite{KasprzakEtAlNature06,Keeling_PRL2006,Yamamoto_RMP,CarusottoCiutiRMP13}. Yet those systems are successfully described by semiclassical theories such as driven-dissipative Gross-Pitaevski equations, leaving open the question about the quantum nature of this phenomenon. More recently the attention has shifted toward strongly correlated quantum lattice models with drive and dissipation, where several works have revealed the existence of limit cycles, i.e. non-stationary solutions of the quantum dynamics for a macroscopic order parameter, at least at the mean field level~\cite{JinetalPRL13,ChanEtAlPRA15,WilsonEtAlPRA16,SchiroPRL16,OwenEtAl_arxiv17,BiellaEtAlPRA17} 

In this work we focus on a paradigmatic model of driven-dissipative interacting bosons on a lattice, which is directly relevant for the upcoming generation of circuit QED arrays experiments targeting Mott insulators of polaritons~\cite{FitzpatrickEtAlPRX17,MaEtAlPRA17,MaArxiv2018}. We argue that a dynamical susceptibility of such an open quantum many body system, which in thermal equilibrium is finite and small since non-zero frequency modes are typically damped by interactions, can display a genuine singularity at finite frequency, as a result of strong interactions and non-equilibrium effects. The critical frequency is not fixed a priori, but rather an emerging scale set by the microscopic parameters. Eventually, the system undergoes a dynamical phase transition where the order parameter in the broken symmetry phase becomes non-stationary and oscillates in time without damping, thus breaking the continuous time-translational symmetry. This stationary-state instability is controlled by both dissipative and coherent couplings, in particular by the ratio between hopping and local interaction, thus providing the strongly correlated analogue of weak coupling non-equilibrium bosons condensation. We organize the manuscript in the following way:
in section~\ref{sec:modelAndDriving} we introduce the many-body model and we discuss the physics of its single-site. In section~\ref{sec:instability} we discuss the stationary-state instability of the normal phase characterized by a diverging finite frequency susceptibility. We present the consequences of this instability on the dissipative evolution in section~\ref{sec:dynamics} and finally, in section~\ref{sec:fieldTheory}, we recover this dynamics from the saddle point solution of the effective non-equilibrium field theory which describes the transition from the strongly correlated regime. In appendix~\ref{app:drive}, we discuss results which are related to the specific pump/loss mechanisms and we consider an alternative driving scheme.

\section{Model and Driving}
\label{sec:modelAndDriving}
We consider the Bose-Hubbard (BH) Hamiltonian 
\be \label{eq:BH}
H=-J\sum_{\langle ij\rangle}\left(a^{\dagger}_ia_j+hc\right)+\sum_i\left(\delta\omega_0 n_i+\frac{U}{2}n_i^2\right)
\ee
modelling a lattice of circuit QED resonators~\cite{raftery_observation_2014,HacohenGourgyPRL15,FitzpatrickEtAlPRX17} hosting a single bosonic cavity mode on each site, $a_i,a^{\dagger}_i$, with frequency $\delta\omega_0$, repulsive local interaction $U$ and hopping rate $J$.
We supplement the model with pump and losses, necessary to drive the system into a non-trivial steady state, and we remark that the qualitative features we are going to discuss do not depend on the specific implementation of drive and dissipation. Here we describe the driven-dissipative dynamics in terms of a master equation for the system density matrix 
\be\label{eqn:master} 
i\partial_t\rho = -i[H,\rho]+\mathcal{D}[\rho]
\ee
where the dissipator takes the form $\mathcal{D}[\rho]=\mathcal{D}_{loss}[\rho]+\mathcal{D}_{pump}[\rho]$. The first term describes single particle losses with rate $ \kappa$ \beq \mathcal{D}_{loss}[\rho]=\kappa\sum_i \left(a_i\rho a_i^{\dagger}-\frac{1}{2}\left\{a^{\dagger}_ia_i,\rho\right\}\right) \eeq while the second term 
\beq\label{eq:incohDiss} \mathcal{D}_{pump}[\rho]=\sum_i f_{in}\tilde{\mathcal{D}}[a^{\dagger}_i,\tilde{a}_{i\sigma};\rho]+f_{out}\tilde{\mathcal{D}}[a_i,\tilde{a}^{\dagger}_{i\sigma};\rho] \eeq
 accounts for the coupling to an incoherent bath with a finite bandwidth $\sigma$, which injects and removes particles without a well defined phase in each cavity, at rates $f_{in},f_{out}$. Here we have introduced the modified dissipator~\cite{Houck_prl11,LebreuillyEtAlPRA17} \beq \tilde{\mathcal{D}}[X,Y]=X\rho Y+Y^{\dagger}\rho X^{\dagger}-X^{\dagger}Y^{\dagger}\rho-\rho YX \eeq and \beq \tilde{a}^{\dagger}_{\sigma}=\sum_n\sqrt{n+1}C^*_{\sigma}\left(\Delta_{n+1}\right)\vert n+1\rangle\langle n\vert\eeq which is the photon operator dressed by the finite-bandwidth drive, with $\Delta_{n+1}=U n+\left(\delta\omega_0+U/2\right)$ the level spacing of an isolated single site in Eq.~\eqref{eq:BH}. 
All the results in the main text are obtained with a microscopic model of pumping describing an ensemble of driven two-level emitters embedded in each cavity and undergoing a population inversion, as recently proposed~\cite{LEBREUILLY2016836,LebreuillyEtAlPRA17,BiellaEtAlPRA17}, which results in $f_{in}=f, f_{out}=0$. We discuss in the the appendix \ref{app:drive} other microscopic models of incoherent pumping~\cite{Houck_prl11} and the effects of changing the driving protocol.
We notice that driving the cavities incoherently preserves the $U(1)$ symmetry, the invariance under
$a_i\rightarrow e^{i\phi} a_i$, of the Hamiltonian in Eq.~(\ref{eq:BH}), 
thus leaving open the possibility for a non-equilibrium phase transition between a Mott-like incoherent phase and a superfluid, which is well understood in equilibrium~\cite{Fisher_Fisher_PRB89,Sachdev}.

\subsection{The single-site problem}
\label{sec:singleSite}
\begin{figure}[t]
\includegraphics[width=7.5cm]{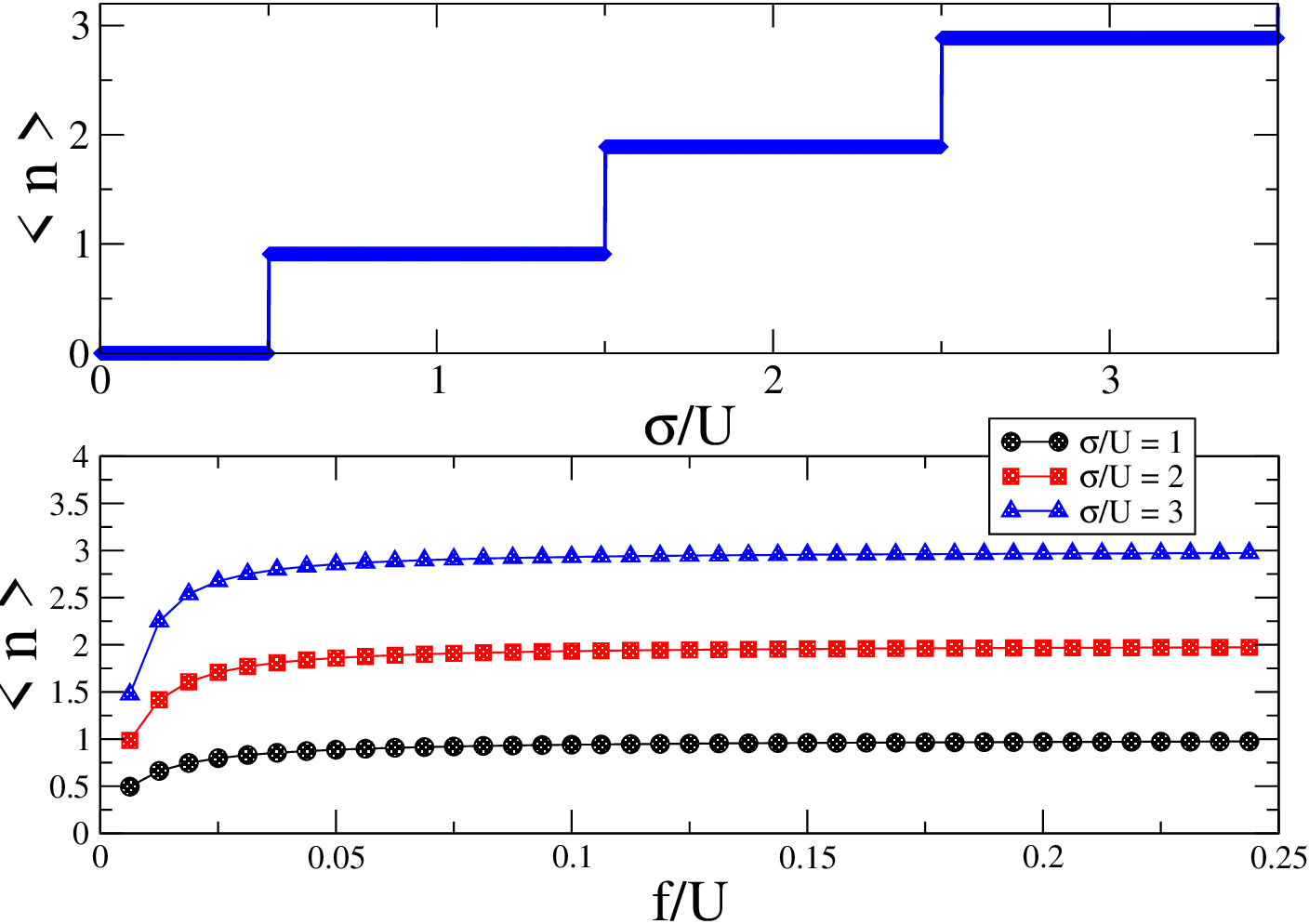}
\caption{Local bosonic occupation in the interacting single-site driven dissipative problem, respectively as a function of the drive bandwidth $\sigma$ (top panel) showing a characteristic staircase structure and drive amplitude $f$ (bottom panel). The value of the local occupation in the plateaux depends strongly on the nature of the driving protocol and can in general be tuned continously with the drive amplitude $f$. Parameters: loss rate $\kappa=0.0128 U$, resonator frequency $\delta\omega_0=0.0$, drive amplitude (top panel) $f=0.0625 U$.  }
\label{fig:fig1}
\end{figure}

To gain some insight on the problem, we start considering the limit $J=0$, where the lattice problem reduces to a collection of interacting, driven-dissipative decoupled sites. While consisting only of a single site, this quantum problem remains quite non-trivial due to non-linearity and dissipation effects and cannot be in general solved analytically, as in the equilibrium case~\cite{Sachdev}. The balance between drive, dissipation and interaction results in a finite number of bosons per site, which however remain incoherent, $\langle a_i\rangle=0$. The boson number in the single-site problem as a function of pump bandwidth $\sigma$, plotted in figure~\ref{fig:fig1} (top panel), shows a starcaise structure characteristic of blockade physics~\cite{Houck_prl11}, with a value of $\Delta\sigma\sim U$ required to add extra bosons in the system.  This can be understood naturally: to add a boson in an interacting site the drive has to provide extra energy, however since the system is ultimately open and boson number is not exactly conserved, the exact occupancy will be fixed by the ratio of pump and losses. The drive is able to fix the occupancy to almost integer filling, reflecting the fact that the stationary density matrix is almost pure and resembling the ground state physics of a Bose-Hubbard interacting site, but we stress that the average density in this open and dissipative implementation is never exactly integer, due to losses~\cite{LebreuillyEtAlPRA17}.  The width of the steps is set by the interaction while the height can be tuned continously by changing pump amplitude $f$, as we show in the bottom panel of figure~\ref{fig:fig1}.

\begin{figure}[t]
\includegraphics[width=\columnwidth]{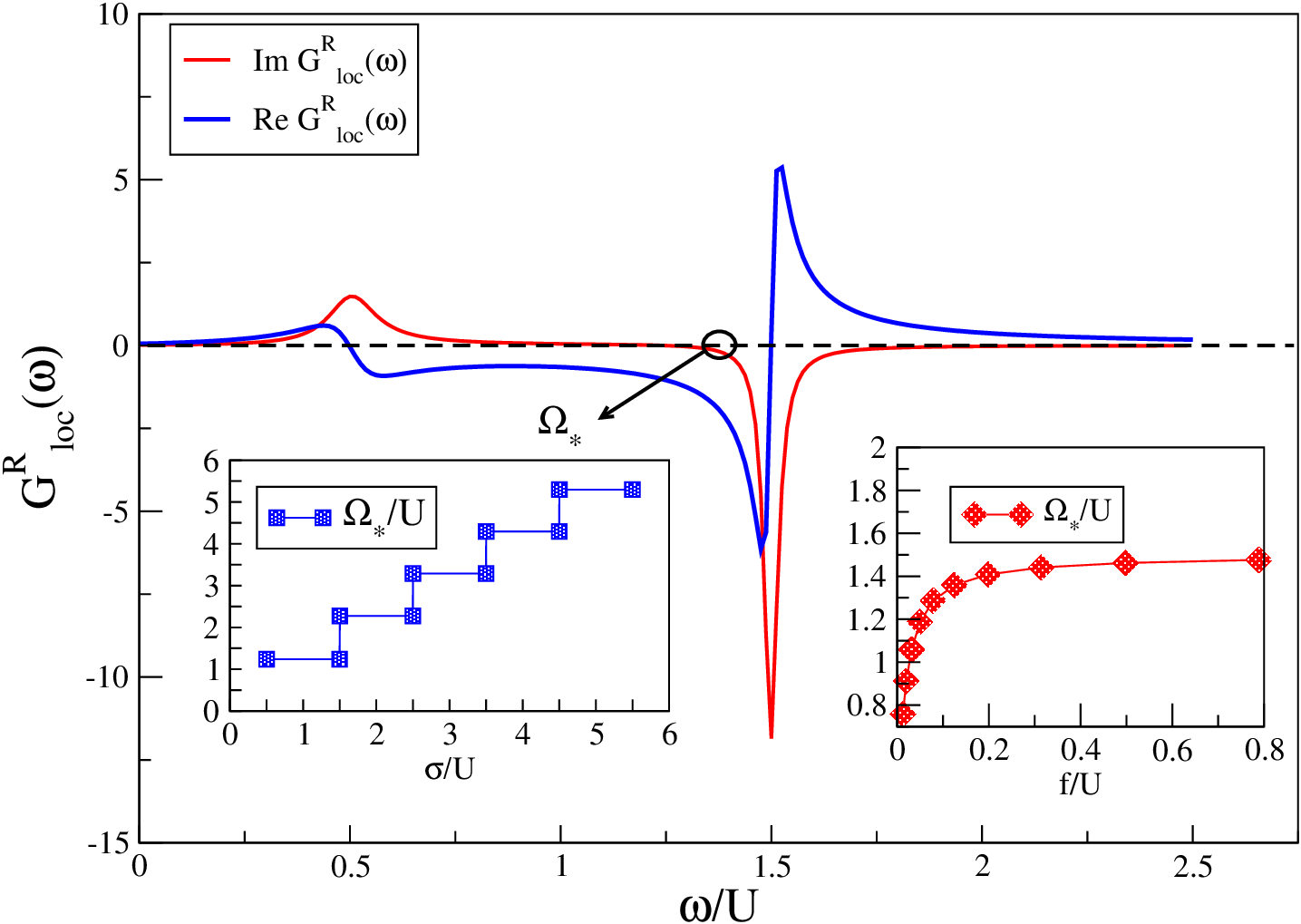}
\caption{Local Spectral Function and the Emergent Frequency $\Omega_*$.  Retarded Green's function, see equation~(\ref{eq:greensSS}) of the interacting single-site problem in Eq.~(\ref{eq:BH}) with drive and dissipation. The imaginary part, describing the spectral function, shows two peaks separated by a gap of order $U$. Crucially, the two peaks are not symmetrically placed around zero frequency, i.e. the imaginary part changes its sign at a finite frequency $\Omega_*$. This emergent frequency is not fixed a priori but rather fully tunable and depending from the amplitude and bandwidth of the drive.  Parameters: Drive amplitude $f=0.0625 U$, bandwidth $\sigma=1.5U$,  loss rate $\kappa=0.0128 U$, resonator frequency $\delta\omega_0=0.0$.}
\label{fig:fig2}
\end{figure}

\begin{figure}[t]
\includegraphics[width=7.5cm]{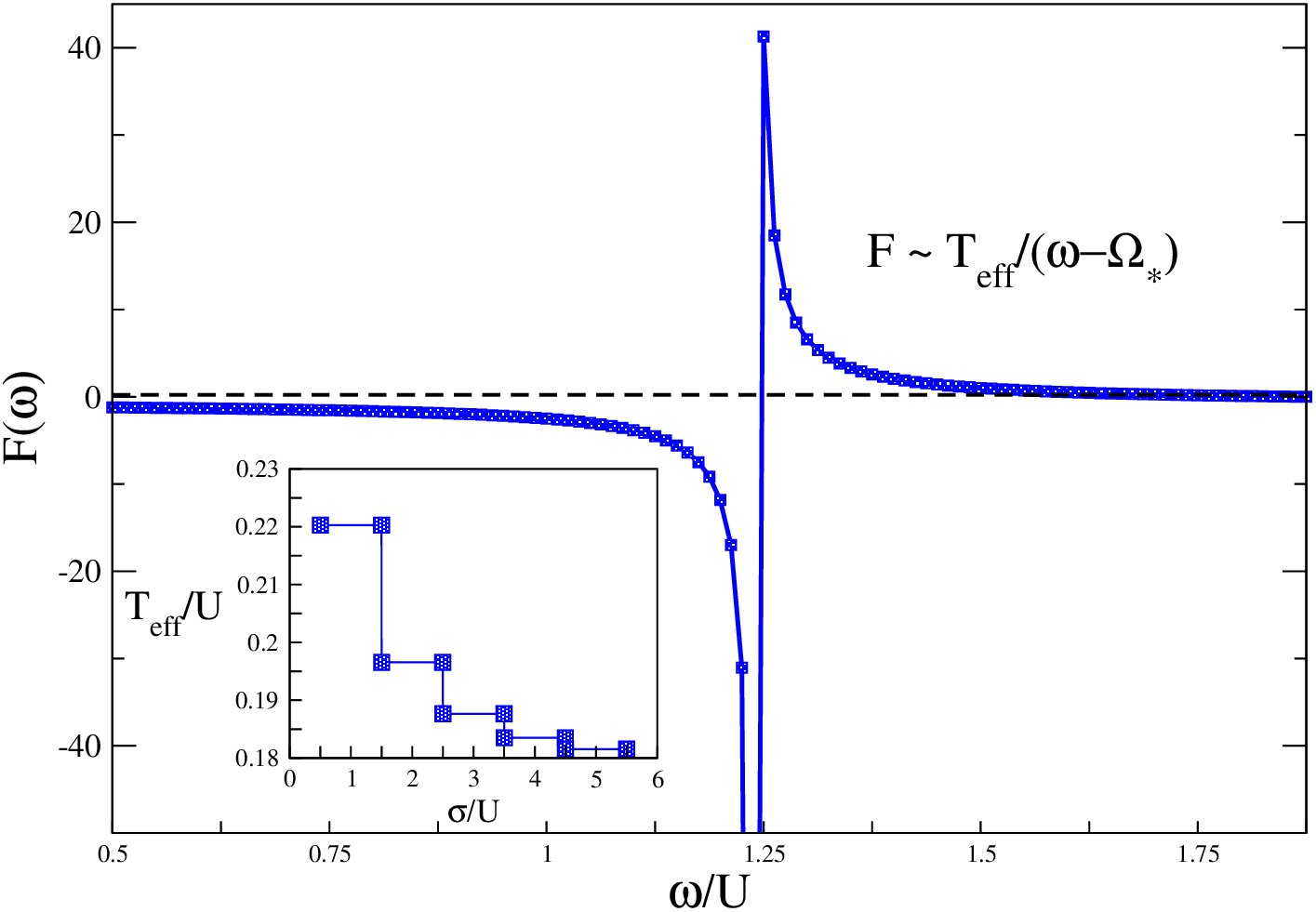}
\caption{Effective Distribution Function as a function of frequency around $\Omega_*$ for fixed value of interaction, drive and dissipation. Around the frequency $\Omega_*$ the system develops a singularity which allows to define an effective temperature $T_{eff}$, whose dependence from the drive bandwidth is plotted in the inset. Parameters: loss rate $\kappa=0.0128 U$, resonator frequency $\delta\omega_0=0.0$, drive bandwidth (left panel) $\sigma= 0.625U$, drive amplitude (right panel) $f=0.0625 U$.   }
\label{fig:fig3}
\end{figure}

A key role in this work is played by the single site Green's functions (retarded/advanced/Keldysh) evaluated in the stationary state which are defined as
\be 
\label{eq:greensSS}
G^{R/K}_{loc}(\omega)= -i\int_0^{\infty} dt e^{i\omega t}\langle [a(t),a^{\dagger}(0)]_{\mp}\rangle_{loc}\,.
\ee
and $G_{loc}^A(\omega) = G_{loc}^R(\omega)^*$, which we evaluate using their Lehmann representation~\cite{ScarlatellaClerkSchiro_arxiv18} and a numerical diagonalization of the Liouvillian.
As we see in the bottom panel of figure~\ref{fig:fig2}, when the interaction $U$ is sufficiently strong with respect to dissipation, the lorentzian-shaped spectrum of a driven-dissipative oscillator splits into two atomic-like excitations separated by a large gap $U$. Remarkably, this spectral function is not constrained to change sign at zero frequency as in thermal equilibrium, but it does it at a non-zero energy scale $\Omega_*\neq 0$ which depends on interactions, drive and dissipation. In figure~\ref{fig:fig2} we plot this quantity as a function of the drive bandwidth $\sigma$ and drive amplitude $f$, showing that  $\Omega_*$ increases with $\sigma$ and $f$, together with the bosonic occupation.
This frequency can be interpreted as an emergent chemical potential for the bosons, as we can deduce by considering the bosonic distribution function, which contains information on the occupation of bosonic modes.
We can define a bosonic distribution function in analogy with the thermal equilibrium case, i.e.
\beq
\label{eqn:FdistSS}
F(\omega) = \frac{G^K_{loc}(\omega)}{G^R_{loc}(\omega) - G^A_{loc}(\omega)}
\eeq
where $G^K_{loc}(\omega)$ is the Keldysh Green's function.
Indeed in thermal equilibrium the fluctuation-dissipation theorem constraints the functional form of the distribution function to the canonical bosonic one, $F_{eq}(\omega)=\coth\beta\omega/2$, which at low frequency (or high temperature) becomes $F_{eq}(\omega)\simeq T/2\omega$. Here, in presence of drive and dissipation, such an identity does not hold and we use Eq.(\ref{eqn:FdistSS}) as operational definition of the distribution function. 
We plot in figure~\ref{fig:fig3} the distribution function for a given value of interaction, drive and dissipation. While its overall shape shows departure from the thermal equilibrium case, we find that around the critical frequency $\Omega_*$ the system develops a singularity of the form $F(\omega)\sim T_{eff}/\left(\omega-\Omega_*\right)$, ultimately arising from the fact that $\mbox{Im}G^R_{loc}(\omega)$ has a zero at $\Omega_*$ while the Keldysh component is finite around the same frequency range. This suggests an asymptotic thermalization around the frequency $\Omega_*$, with a small effective temperature $T_{eff}$ weakly decreasing with the drive bandwidth (see inset of figure~\ref{fig:fig3}) and with the frequency $\Omega_*$ playing now the role of an effective chemical potential. 


\section{Instability of Normal Phase}
\label{sec:instability}
We now turn to discuss the properties of the lattice model in presence of a finite hopping $J$. To this extent we introduce a general approach to study the instabilities of the normal incoherent phase of driven-dissipative correlated lattice models, which generalizes the equilibrium strong-coupling approach of Ref~\cite{Fisher_Fisher_PRB89}. We start writing the Keldysh action associated to the many-body quantum master equation~(\ref{eqn:master}),  which allows to describe the non-equilibrium stationary state and the excitations on top of it, and decouple the hopping term by means of an auxiliary bosonic field playing the role of local order parameter. The resulting effective action takes the form
\be \label{eqn:Seff}
\mathcal{S}_{eff}=\int_C dt\sum_{ij}\psi_i^{\dagger}J_{ij}^{-1}\psi_j+\sum_i\Gamma[\psi^{\dagger}_i,\psi_i]
\ee 
 where the second term represents the generating functional of the local bosonic Green's functions, $\Gamma[\psi_i^{\dagger},\psi_i]=\log\langle T_C e^{i\int_C dt\left(\psi^{\dagger}_ia_i+a^{\dagger}_i\psi_i\right)}\rangle_{loc}$, with the average taken over the interacting driven-dissipative single-site problem. Our approach therefore combines the strong coupling field theory~\cite{Fisher_Fisher_PRB89} with the exact numerical solution of the single-site problem based on a diagonalization of its Liouvillian. As such, it could be  applied to lattice models with any scheme of incoherent local drive and dissipation by just solving the appropriate local problem.

Expanding Eq.~(\ref{eqn:Seff}) in the fields $\psi,\psi^{\dagger}$ and within a gaussian approximation, we obtain the effective action $\mathcal{S}_{eff}=\int d\omega d^dq \bar{\Psi}(q,\omega) \mathcal{\hat{G}}^{-1}(q,\omega)\Psi(q,\omega)$, where we moved to momentum space, we defined $\bar{\Psi}(q,\omega)= \left(\psi^*_c(q,\omega)\; \psi_q^*(q,\omega)\right)$ with $\psi_{c/q}$ the classical/quantum fields, and we introduced
$$\label{eqn:Gquad}\label{eqn:Ginv_qomega}
\mathcal{\hat{G}}^{-1}(q,\omega)= 
\left(
\begin{array}{ll}
0 & J_q^{-1}-G_{loc}^A(\omega)\\
J_{q}^{-1}-G_{loc}^R(\omega) & -G_{loc}^K(\omega)
\end{array}
\right)
$$
In the above expression we have  $J_q=-2J\sum_{\alpha=1}^d\cos q_{\alpha}$, while $G^{R/A/K}_{loc}(\omega)$ are the exact single-site retarded/advanced/Keldysh Green's functions evaluated in the stationary state

From the effective action $\mathcal{S}_{eff}$, the susceptibility $\chi^R(q,\omega)$ of the order parameter reads 
\be \label{eqn:suscept}
\chi^R(q,\omega)=\frac{1}{J_q^{-1}-G^R_{loc}(\omega)}
\ee 
\begin{figure}[t]
\includegraphics[width=\columnwidth]{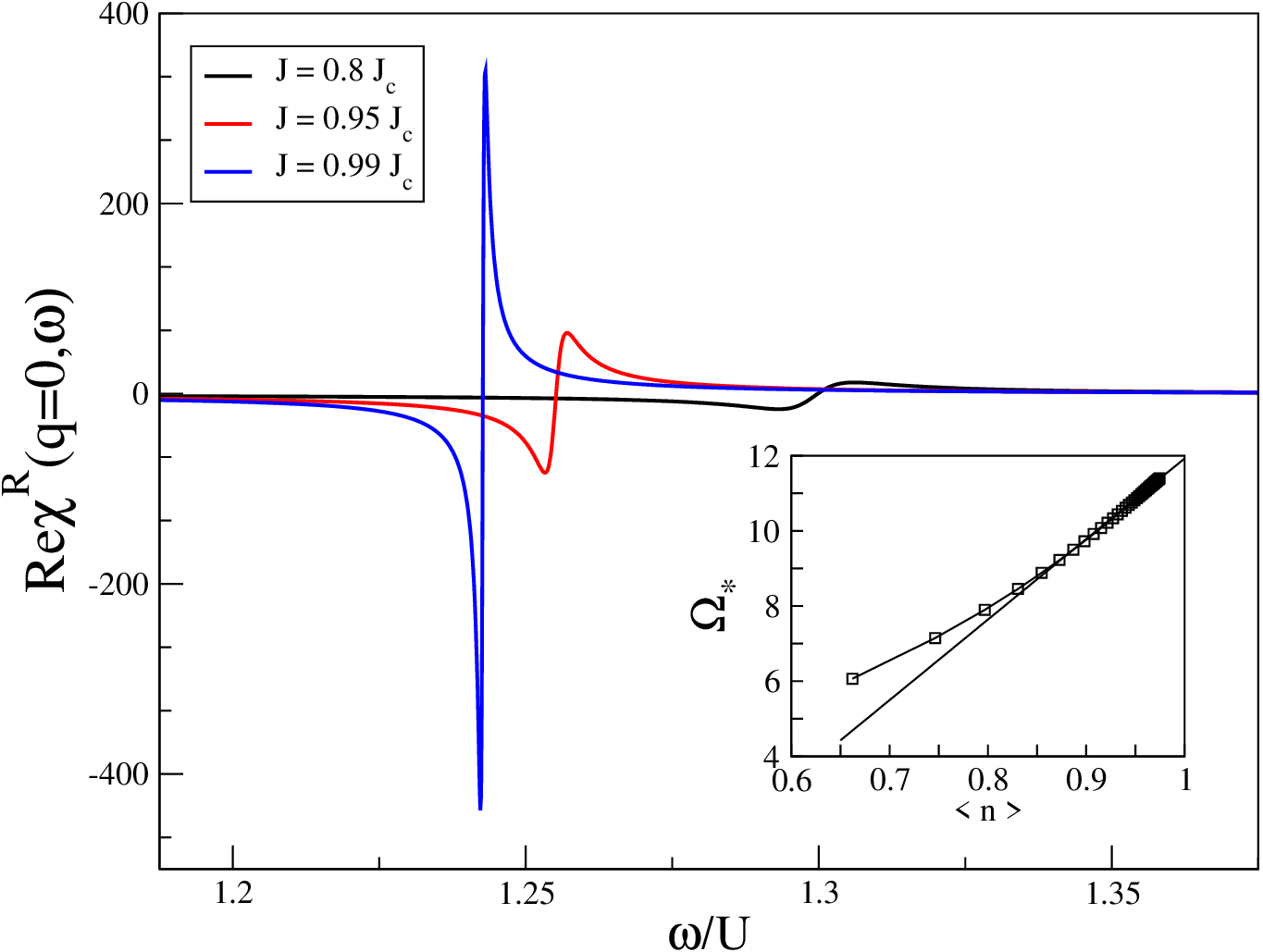}
\caption{Normal phase susceptibility in the gaussian approximation, Eq.~(\ref{eqn:suscept}), as a function of frequency and for different values of the hopping $J$. The real part of the susceptibility goes through a zero with a characteristic two-peaks structure which gets sharper as the critical value $J_c$ is approached, turning into a genuine singularity at $\Omega_*$. This emergent scale is controlled by the local occupation of the single site problem (top panel), whose spectral function is plotted in the bottom inset. Parameters: Drive amplitude $f=0.0625 U$, bandwidth $\sigma=1.5U$, loss rate $\kappa=0.0064 U$, resonator frequency $\delta\omega_0=0.0$.}
\label{fig:fig4}
\end{figure}
The gaussian approximation of the effective action, leading to Eq.~(\ref{eqn:suscept}), is well suited within the normal phase, where the order parameter fluctuates around zero. It is equivalent to a strong-coupling resummation of the perturbation theory in the hopping $J$ often referred to as random phase approximation (RPA) around the atomic limit~\cite{SchmidtBlatter_PRL09,Koch_LeHur_2009} and it is well known to capture qualitatively the instability of the normal phase, as for example discussed in Refs.~\cite{Fisher_Fisher_PRB89,Sachdev}.

In thermal equilibrium, the $U(1)$ susceptibility Eq.~(\ref{eqn:suscept}) is well known to show a zero-frequency singularity at a critical value of the hopping, at which the Mott insulating phase becomes unstable towards superfluidity~\cite{Fisher_Fisher_PRB89,Sachdev}. As we are going to show, the behavior of the same quantity in a non-equilibrium state is remarkably different. 
In figure~\ref{fig:fig4}  we plot the $q=0$ susceptibility, probing the instability of the homogeneous normal phase,  for different values of the hopping strength $J$. We find a well defined resonance structure which gets sharper and narrower as the hopping is increased, and eventually turns into a genuine \emph{finite frequency} pole at $\omega=\Omega_*$ when a critical hopping $J_c$ is reached. Right at $J_c$ the susceptibility diverges as a power law around $\Omega_*$,   $\chi^R(q=0,\omega)=\chi_0/\left(\omega-\Omega_*\right)^{\alpha}$, with $\alpha=1$. 
The appearance of a singularity at finite frequency is a remarkable result with no counterpart in systems in thermal equilibrium,  where one expects finite frequency modes to be damped by interactions thus acquiring a finite lifetime, ultimately cutting off the singularity of any dynamical susceptibility.  
Its origin is rooted in the physics of the single-site quantum problem whose spectral function, $\im G_{loc}^R(\omega)$, enters the  susceptibility through Eq.~(\ref{eqn:suscept}). In particular, as we show in the top panel of figure~\ref{fig:fig4}, the critical frequency is set by the local occupancy, rather than by the strenght of the order parameter as for the oscillations of weakly interacting non-equilibrium superfluids~\cite{DunnetEtAlPRB16}, a fact which highlights the quantum nature of the incoherent phase becoming unstable at $\Omega_*$.

%

\begin{figure}[t]
\includegraphics[width=\columnwidth]{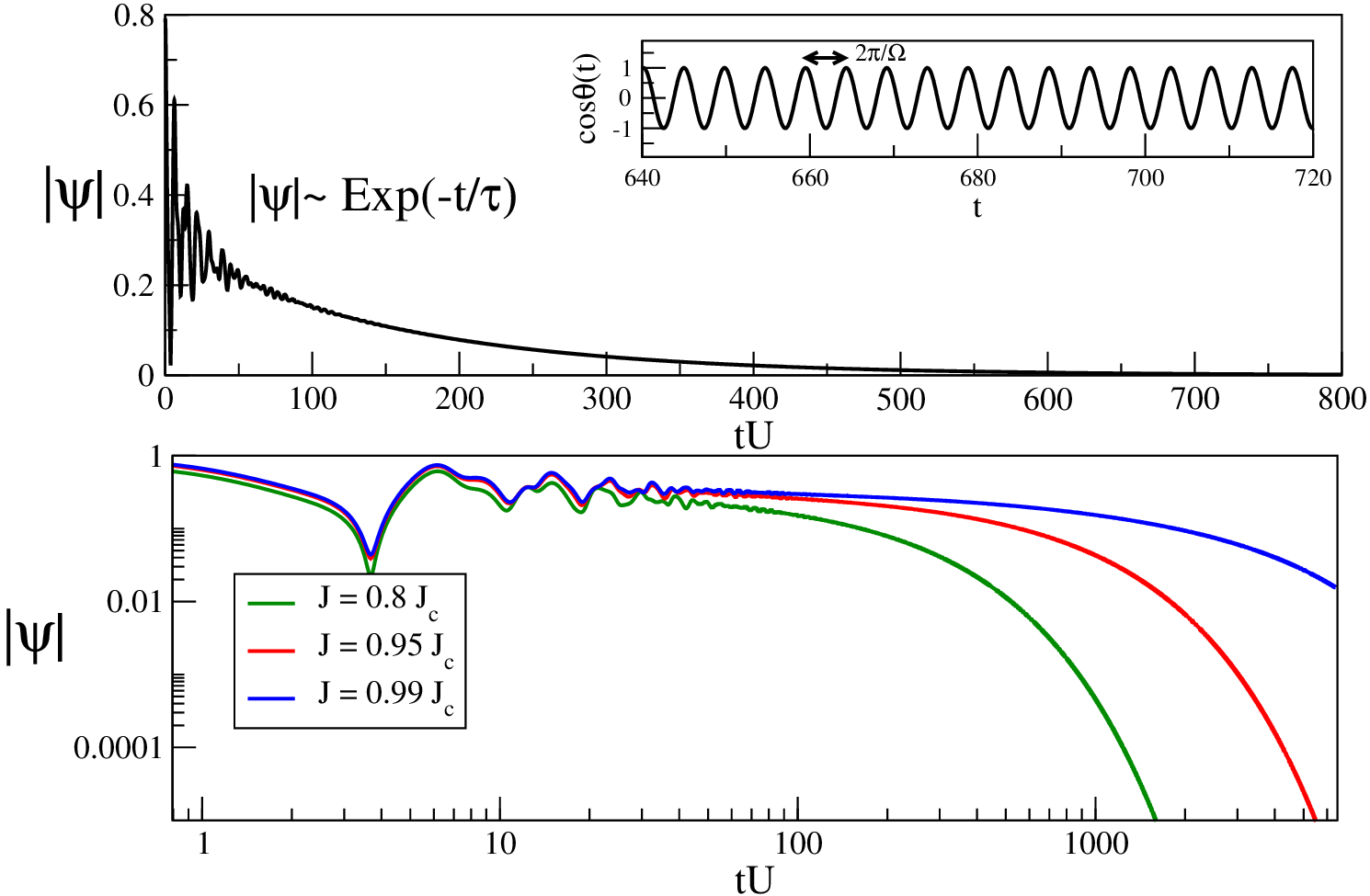}
\caption{Dynamics of the order parameter $\psi(t)=\vert\psi(t)\vert e^{i \theta(t)}$ obtained from the tdG method, for different values of the hopping strength $J$. (Top Panel) Normal phase, $J=0.8J_c$, exponential decay of the order parameter toward an incoherent stationary state $\vert\psi(t\rightarrow\infty)\vert\sim e^{-t/\tau(J)}$, with oscillations at frequency $\Omega(J)$ (see inset) due to the phase  linearly growing in time. (Bottom Panel, log scale) Approaching the critical point $J_c$, the dynamics slows down suggesting a power-law decay right at the transition, as we show analytically. Parameters: Drive amplitude $f=0.0625 U$, bandwidth $\sigma=1.5U$, loss rate $\kappa=0.0064 U$, resonator frequency $\delta\omega_0=0.0$.} 
\label{fig:fig5}
\end{figure}
\section{Dissipative Dynamics}
\label{sec:dynamics}
We now discuss the consequences of the finite-frequency, stationary-state instability we have presented so far and investigate the dissipative dynamics~(\ref{eqn:master}) of the lattice problem for different values of the hopping $J$. To this extent, we use a time-dependent Gutzwiller (tdG) decoupling of the density matrix, i.e. $\rho(t)=\prod_i \rho_i(t)$ that we further assume homogeneous in space, $\rho_i(t)\equiv \rho_{loc}(t)$.  This approximation results in an effective single-site problem $\partial_t\rho_{loc}(t)= -i[H_{eff}(t),\rho_{loc}(t)]+\mathcal{D}_{loc}[\rho_{loc}]$ where $H_{eff}(t)=\delta\omega_0 n +Un^2/2+zJ\left(a^{\dagger}\psi(t)+hc\right)$ with $z=d/2$ the coordination number of the lattice, $\mathcal{D}_{loc} $ is the local dissipator including incoherent drive and losses, while $\psi(t)=\mbox{Tr}\rho_{loc}(t)a$ is a self-consistent time dependent field.  
We expect this approximation to capture some qualitative features of the dynamics across the phase transition, at least in high enough dimensions, where its mean field description is supposed to be accurate.

\begin{figure}[t]
\includegraphics[width=\columnwidth]{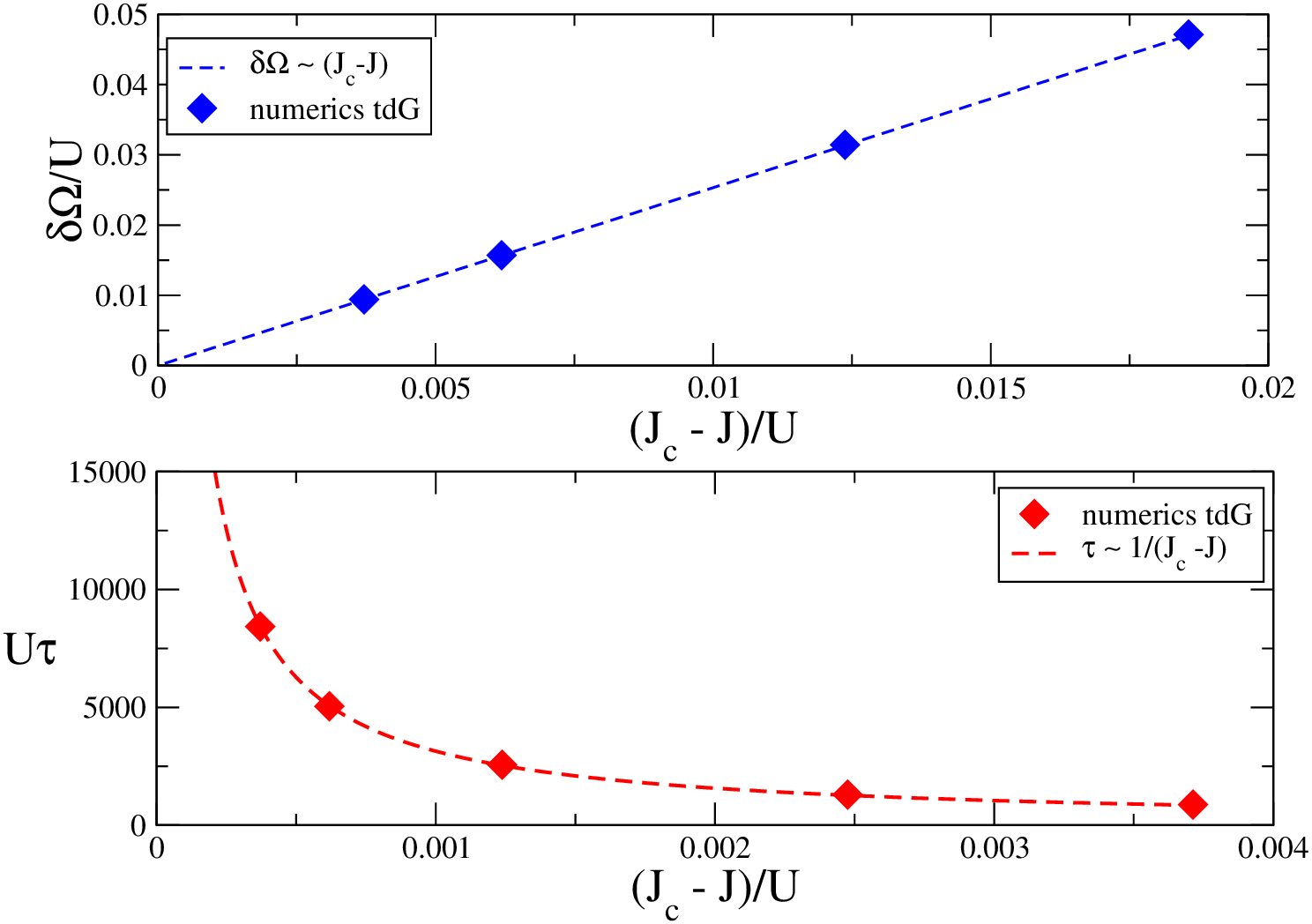}
\caption{Comparison of the results obtained from the tdG method and from Eq. \eqref{eqn:saddle_point} for the two scales $\delta\Omega$ (slow frequency oscillation mode, top panel) and $\tau$ (relaxation time to reach a steady state, bottom) as a function of the distance from the critical point, $J_c-J$. The dissipative dynamical transition is characterized by both energy scales becoming critical. Parameters: Drive amplitude $f=0.0625 U$, bandwidth $\sigma=1.5U$, loss rate $\kappa=0.00648 U$, resonator frequency $\delta\omega_0=0.0$.}
\label{fig:fig6}
\end{figure}

In figure~\ref{fig:fig5} we plot the dynamics of the bosonic order parameter $\psi(t)=\langle a(t)\rangle$ as a function of time  for different values of the hopping $J<J_c$. If we introduce a polar representation, $\psi(t)=\vert\psi(t)\vert e^{i \theta(t)}$, we see that the absolute value of the order parameter shows an exponential relaxation toward zero, $\vert\psi(t)\vert\sim e^{-t/\tau(J)}$, indicating an incoherent stationary state, while the phase grows linearly in time with finite angular velocity $\Omega$,  $\theta(t)=\Omega(J) t+\theta_0$.
A closer inspection reveals that the characteristic frequency $\Omega(J)$ differs from the value $\Omega_*$ previously identified by an amount $\delta\Omega(J)=\vert \Omega(J)-\Omega_*\vert$ which strongly depends on the hopping rate $J$ and vanishes at the critical point $J=J_c$ with a characteristic power law, $\delta\Omega\sim (J_c-J)$, as shown in the top panel of figure~\ref{fig:fig6}. Similarly the relaxation time diverges upon approaching the critical hopping $J_c$, $\tau\sim 1/(J_c-J)$ (see figure~\ref{fig:fig6}) and the order parameter shows a characteristic critical slowing down, as shown in the bottom panel of figure~\ref{fig:fig5}. 

\section{Nonequilibrium Field Theory of Finite Frequency Criticality}
\label{sec:fieldTheory}
 We now proceed to set up a Keldysh non-equilibrium field theory for the finite-frequency dissipative transition, which allows us to obtain a complete analytical picture of the mean field dynamics and sets the stage to discuss the role of quantum fluctuations beyond mean field. The starting point is to expand the effective action~(\ref{eqn:Seff})  for $q\rightarrow0 $ and  $\omega\rightarrow\Omega_*$ and then move to a rotating frame where the field is oscillating at frequency $\Omega_*$. Introducing the fields $\tilde{\psi}_{c,q}(x,t)=e^{-i\Omega_*t}\psi_{c,q}(x,t)$, we obtain
\bea\label{eqn:Seff_fieldtheory}
\mathcal{S}_{eff}&=&\int dt dx \tilde{\psi}_c^*\left(-r+K_1 i\partial_t+\frac{K_2}{2}\partial_t^2-K_3\nabla^2\right)\tilde{\psi}_q+ hc\nonumber\\
&&+ \mathcal{S}_{noise}+\mathcal{S}_{int}
\eea
where $r=1/zJ+\mbox{Re}G^R_{loc}(\Omega_*)=\left(J_c-J\right)/J_c^2 $ is the distance from the dissipative phase transition while  $K_3=1/zJ^2$ is the bare mass. Differently from the Gross-Pitaevski weak-coupling regime~\cite{SiebererRepProgPhys2016}, the effective action~(\ref{eqn:Seff_fieldtheory}) features both first and second time-derivative terms, with complex coefficients $K_{1,2}=\partial^{1,2}_{\omega}G^R_{loc}(\Omega_*)$, a feature of the strong-coupling limit around which we expand.  In equilibrium $K_1,K_2$ play a crucial role for the critical behavior of the transition, which changes universality class at the tip of the Mott lobes, where $K_1=0$. In the present case we find $K_1\neq 0$ along the phase boundary, suggesting a single universality class for our driven-dissipative case. We therefore drop $K_2$ in the following, an assumption which is justified at the mean field level, and leave for future studies the investigation on the role of $K_2$ upon including fluctuations. In Eq.~(\ref{eqn:Seff_fieldtheory}), $\mathcal{S}_{noise}=\int dxdtdt'\, \tilde{\psi}_q^*(x,t)G^K_{loc}(t-t')\tilde{\psi}_q(x,t')$ represents the noise contribution, which depends on the Keldysh Green's function of the single site. Around $\Omega_*$ this is finite which suggests to disregard retardation and obtain a purely time local quadratic action with noise term $\mathcal{S}_{noise}=D\int dxdt \tilde{\psi}_q^*(x,t)\tilde{\psi}_q(x,t)$. Here $D$ plays the role of effective diffusion coefficient in the equivalent stochastic (Langevin) dynamics and it is indeed given by $D\simeq G^K_{loc}(\Omega_*)\sim T_{eff}$.  Finally $\mathcal{S}_{int}$ accounts for the non-linearities and it is completely determined by the multi-particle Green's functions of the driven-dissipative single site problem.  If we restrict ourselves to interaction terms with one quantum and three classical fields, which is valid in high enough dimensions according to canonical power counting~\cite{SiebererRepProgPhys2016}, we can write this term as
$\mathcal{S}_{int}=u\int dx dt\tilde{\psi}^*_q(x,t)\tilde{\psi}^*_c(x,t) \tilde{\psi}^2_c(x,t) +hc$. We can now take the saddle point equation $\delta S/\delta\tilde{\psi}^*_q(x,t)=0$ and obtain the equation of motion 
\be \label{eqn:saddle_point}
\left(iK_1\partial_t-K_3\nabla^2-r\right)\tilde{\psi}_c+u\vert\tilde{\psi}_c\vert^2\tilde{\psi}_c =0
\ee
which takes the form of a complex Ginzburg-Landau equation, well known as a phenomenological description of pattern formation in classical non-equilibrium systems~\cite{CrossHohenbergRMP93,CrossGreenside_Pattern,VanSaarloosPhyRep03}. 
The spatially homogeneous solution of Equation~(\ref{eqn:saddle_point}) can be obtained in closed form, as we discuss in the next section, and it describes a transition between a phase where $\tilde{\psi_c}\rightarrow 0$ for $t\rightarrow \infty$ and a phase where the modulus of the order parameter saturates to a finite value.
It further allows us to compute the scales $1/\tau$ and  $\delta\Omega$ which perfectly match the numerical results that we have found from the Gutzwiller dynamics in the previous section, as shown in Figure~\ref{fig:fig5}. 
We remark that this classical dynamics only describes the evolution in a frame rotating at frequency $\Omega_*$. The origin of this energy scale, which is not contained in Eq.~(\ref{eqn:saddle_point}), is instead genuinely quantum mechanical and rooted in the solution of the quantum single site problem, as previously discussed.

\subsection{Analytical Solution of Saddle Point Dynamics}

We discuss here more in detail the solution of Equation~(\ref{eqn:saddle_point}). Assuming an homogeneous solution and introducing polar coordinates for absolute value and phase of the order parameters $\tilde{\psi}_c(t)\equiv \vert\tilde{\psi}_c(t)\vert e^{i \tilde{\theta}(t)}$ one obtains two separate equations for $\vert\tilde{\psi}_c(t)\vert$ and $\tilde{\theta}(t)$ 
whose solution reads
\bea\label{eqn:abs}
\vert\tilde{\psi}_c(t)\vert = \vert \tilde{\psi}_c(0)\vert 
\frac{e^{-\tilde{r}_I t}}
{\sqrt{1+\alpha\left(1-e^{-2\tilde{r}_I t}\right)}}\\
\label{eqn:phase}
\tilde{\theta}(t)=-\tilde{r}_R t+\tilde{u}_R\int_0^t dt' \vert\tilde{\psi}_c(t')\vert^2
\eea
where $\tilde{r},\tilde{u}$ are complex coefficients given by 
\bea
\tilde{r}\equiv \tilde{r}_R+i\tilde{r}_I=r/K_1\\
 \tilde{u}\equiv \tilde{u}_R+i\tilde{u}_I=u/K_1
\eea
 while $\alpha=\vert\tilde{\psi}_c(0)\vert \vert \tilde{u}_I\vert/\tilde{r}_I$.   This solution describes a dynamical transition at a critical point $J_c$. Specifically for $J<J_c$ the order parameter shows damped oscillation toward zero 
\be 
\tilde{\psi}_c(t)\sim e^{- t/\tau}\,e^{-i\delta\Omega t}
\ee
with a divergent relaxation time $\tau\sim 1/\tilde{r}_{I}\sim 1/\left(J_c-J\right)$ and an oscillation frequency going to zero $\delta\Omega\sim \tilde{r}_R\sim (J_c-J)$. Upon crossing the critical point, for  $J>J_c$, the dynamics shows instead an amplification of the order parameter which saturates at long time into a train of finite amplitude oscillations 
\be 
\tilde{\psi}_c(t) \sim \vert\tilde{\psi}_c\left(\infty\right)\vert e^{-i\delta\Omega_{sf} t}
\ee
with $\vert \tilde{\psi}_c\left(\infty\right)\vert=\sqrt{\vert \tilde{r}_I\vert /\vert \tilde{u}_I\vert} \sim \sqrt{J-J_c}$ and 
$\delta\Omega_{sf}\sim \left(\tilde{r}_R+\tilde{r}_I\tilde{u}_R/\tilde{u}_I\right)\sim \left(J-J_c\right)$. In the normal phase, where the non-linearity $u$ in Eq.~(\ref{eqn:saddle_point}) is essentially irrelevant, the transient dynamics shows harmonic oscillations while in the broken symmetry phase multiple frequencies are present, at least on intermediate time scales, as encoded in the phase dynamics~(\ref{eqn:phase}). Right at the transition, for $J=J_c$ when $\tilde{r}_{R,I}\rightarrow 0$, the amplitude of the order parameter decays towards zero as a power-law~\cite{Tomadin_prl10,Tomadin_PRA11} while the angular velocity vanishes and the phase grows in time only logarithmically,
\bea
\vert\tilde{\psi}_c(t)\vert\sim 1/\sqrt{t}\\
\tilde{\theta}(t)\sim \log(1+2\tilde{u}_I\vert\tilde{\psi}_c(0)\vert t)
\eea

\subsection{Discussion}

In the previous section we have shown that disregarding (i) retardation effects in the effective action, i.e. expanding all local correlators around the critical frequency $\Omega_*$, as well as (ii) disregarding terms higher than quadratic in the quantum fields allow to fully reproduce the results obtained by time-dependent Gutzwiller decoupling, describing the finite-frequency dynamical transition at the mean field level. Still the full effective action in Eq.~(\ref{eqn:Seff_fieldtheory}) includes the effect of non linearities, noise and quantum fluctuations beyond this semiclassical mean field dynamics and can result in a renormalization of critical behavior and non-mean field exponents~\cite{DeDominicisEtAlPRB75,HohenbergHalperinRMP77,SiebererRepProgPhys2016}. 
These can be captured with a renormalization group treatment of the finite-frequency criticality, along the lines discussed for the equilibrium Bose Hubbard model~\cite{RanconDupuisPRB11} as well as for weakly interacting non-equilibrium superfluids~\cite{SiebereHuberAltmanDiehlPRL13,MarinoDiehlPRB16}. Particularly interesting in this respect is the role under renormalization of higher order expansion coefficients in the effective action, such as $K_1,K_2$ and the next order noise term controlled by $\partial_{\omega}G^K_{loc}(\Omega_*)$. The former are related to emergent symmetries, such as particle-hole in the equilibrium Bose-Hubbard model~\cite{Sachdev} or the asymptotic equilibrium symmetry in the driven-dissipative condensation, whose deviation results in a KPZ like critical phase dynamics~\cite{SiebererRepProgPhys2016} . A term analogous to the latter was shown instead to give rise to non-trivial critical behavior in diffusively driven one dimensional bosons~\cite{MarinoDiehlPRL16}.
An interesting question is whether the inclusion of fluctuations beyond the gaussian level could completely wash away the finite-frequency transition or renormalize the critical frequency $\Omega_*$ down to zero, resulting in a static transition.  While answering this question certainly deserves further investigation our results suggest that, provided the effective action~(\ref{eqn:Seff_fieldtheory}) in the rotating frame admits a non-vanishing $U(1)$ order parameter $\tilde{\psi}_c\neq0$, which is expected in high enough dimensions, then the broken symmetry phase in the original frame will display undamped oscillations and breaking of time-translational invariance.

\section{Conclusions}

 In this work we have shown that a prototype model of correlated driven-dissipative lattice bosons develops, for a critical value of the hopping rate, a diverging susceptibility at a non-zero frequency $\Omega_*$. The resulting finite-frequency criticality corresponds to the dissipative dynamics lacking of a stationary state and rather oscillating in time without damping. Writing down the effective Keldysh field theory for this finite frequency transition we have obtained its semiclassical limit which we show to reproduce the results of a time-dependent Gutzwiller decoupling of the density matrix.  
 We emphasize that knowing the critical frequency $\Omega_*$ requires the quantum solution of the single-site dissipative interacting problem and it is therefore not contained in the semi-classical equation of motion which only describes the dynamics in the frame rotating at $\Omega_*$.  Our results differ from other studies of limit cycles instabilities in driven-dissipative systems, such as exciton-polariton condensates described by Gross-Pitaevski (GP) types of equation and it could be seen as the strongly correlated version of them. Indeed our transition shares genuine features of a dissipative Mott-Superfluid quantum phase transitions being tuned both by coherent couplings and pump/loss rates. In particular our incoherent phase exists at small hopping even beyond the standard threshold of pump greater than losses, an effect which is genuinely quantum mechanical due to the Hubbard repulsion favouring Fock-like states rather than coherent states. Furthermore the frequency of the limit cycle is set by the local occupation rather than the local coherence as in GP theories.  
 
Our work suggests several interesting future directions. From one side it would be interesting to include dynamical and spatial fluctuations on top of the semiclassical dynamics for the order parameter and study the fate of this dissipative dynamical transition in finite dimensions, following similar investigations done for dynamical transitions in isolated quantum systems~\cite{SciollaBiroli_PRB13,MaragaChiocchettaMitraGambassiPRE15,ChiocchettaEtAlPRB16}. Another intriguing open question is whether a similar finite-frequency criticality exists in models of driven-dissipative systems with discrete broken symmetry phases~\cite{SchiroPRL16} or even in presence of a purely coherent drive, as for example in the context of optomechanical platforms~\cite{LudwigMarquardtPRL13,LevitanEtAlNJP16} or coherently driven quantum spin chains~\cite{ChanEtAlPRA15}.
 
 Finally, while our work focuses on a paradigmatic model of driven-dissipative bosons which is relevant for the upcoming generation of circuit QED arrays experiments~\cite{FitzpatrickEtAlPRX17,MaEtAlPRA17}, it  also outlines a generic framework to study dynamical instabilities in non-equilibrium quantum systems, by focusing on frequency dependent response functions and their divergences.  Such a framework has the potential to be applied in a wide range of contexts, including for example driven and isolated Floquet systems, where breaking of discrete time-translational symmetry has been predicted~\cite{KhemaniEtAlPRL16,ElseBauerNayakPRL16,VonKeyserlingkEtAlPRB16} and observed~\cite{ZhangEtAlNature17,ChoiEtAlNature17}, quantum systems undergoing various forms of synchronization~\cite{LudwigMarquardtPRL13,WitthautEtAlNatComm17,LorchEtAlPRL17} as well as electronic systems under pump-probe optical-irradiation~\cite{NagEtAlArxiv18}
 
%
%

%
%

\emph{Acknowledgements.} We acknowledge discussions with A. Clerk, M. Goldstein, V. Savona. This work was supported by the CNRS through the PICS-USA-147504, by a grant "Investissements d'Avenir" from LabEx PALM (ANR-10-LABX-0039-PALM) and by a grant IRS-IQUPS of University Paris-Saclay.

\appendix

\section{Role of Driving Protocol}
\label{app:drive}
The results we have described concerning the finite frequency criticality are extremely robust with respect to the specific driving protocol, provided that bosons are injected in the lattice incoherently. There are however interesting differences in the nature of the normal phase which strongly depend on the nature of the drive, as we are going to discuss below. 
In this work we have considered two driving protocols which we discuss more in detail here and we address the main qualitative differences in the results for the two cases.

In the main text, we considered a scheme of incoherent pumping, which has been recently proposed~\cite{LEBREUILLY2016836,LebreuillyEtAlPRA17,BiellaEtAlPRA17}, arising from an ensemble of $N_{at}$ driven two-level emitters embedded in each cavity and having randomly distributed frequencies. In the following we will refer to this scheme as \emph{cold drive}, for reasons which will become clear in the next sections. The microscopic Hamiltonian for this driving scheme reads
\be\label{eqn:pump2}
H^{cold}_{pump}=\sum_i \sum_{n=1}^{N_{at}} \omega_{at}^{(n)}\sigma^{+(n)}_i\sigma_i^{-(n)}+ g\sum_{i,n}\left(a^{\dagger}_i\sigma^{-(n)}_{i}+hc\right)\,
\ee
where the transition frequencies $\omega_{at}^{(n)}$ of the two-level systems are assumed to be uniformly distributed over a finite range and each emitter is incoherently pumped in the excited state~\cite{LebreuillyEtAlPRA17}. 

In this appendix we consider also a second scheme~\cite{Houck_prl11}, which we will refer to as \emph{hot drive} in the following, where a random classical drive is modulated with a coherent tone as described by the time-dependent Hamiltonian 
\be\label{eqn:pump1}
H^{hot}_{pump}(t)=\sum_i\left(e^{i\omega_L t}a^{\dagger}_i\eta(t)+hc\right)
\ee 
where $\eta(t)$ is assumed to have gaussian statistics with zero average,  $\langle \eta(t)\rangle=0$, and correlations $\langle\eta(t)\eta(t')\rangle=fC_{\sigma}(t-t')$. We assume the noise spectrum to be box-shaped with a finite bandwidth $\sigma$, i.e. $C_{\sigma}(\omega)=\theta(\sigma-\vert\omega\vert)$, and amplitude $f$, although the results we obtain do not depend qualitatively from the exact shape of $C_{\sigma}(\omega)$.
Treating the incoherent driving at the master-equation level we obtain, in both cases of Eqs.~(\ref{eqn:pump1}-\ref{eqn:pump2}), the contribution to the dissipator in Eq.~\eqref{eq:incohDiss} which we report here for clarity~\cite{Houck_prl11,LebreuillyEtAlPRA17}
\be 
\mathcal{D}_{pump}[\rho]=\sum_i f_{in}\tilde{\mathcal{D}}[a^{\dagger}_i,\tilde{a}_{i\sigma};\rho]+f_{out}\tilde{\mathcal{D}}[a_i,\tilde{a}^{\dagger}_{i\sigma};\rho]
\ee 
where we have already introduced the modified dissipator $\tilde{\mathcal{D}}$ and the photon operator dressed by the finite bandwidth drive $\tilde{a}^{\dagger}_{\sigma}$ in section \ref{sec:modelAndDriving}. In the random noise case, Eq.~(\ref{eqn:pump1}), we obtain $f_{in}=f_{out}\equiv f$, namely the drive 
acts both as a source and as a sink of particles, much like a finite-temperature bath. Instead, in the case of inverted random emitters we have $f_{in}=f$ and $f_{out}=0$, namely there are no additional losses of particles associated to the drive.

\begin{figure}[t]
\label{fig:tempVsSigma}
\includegraphics[width=7.5cm]{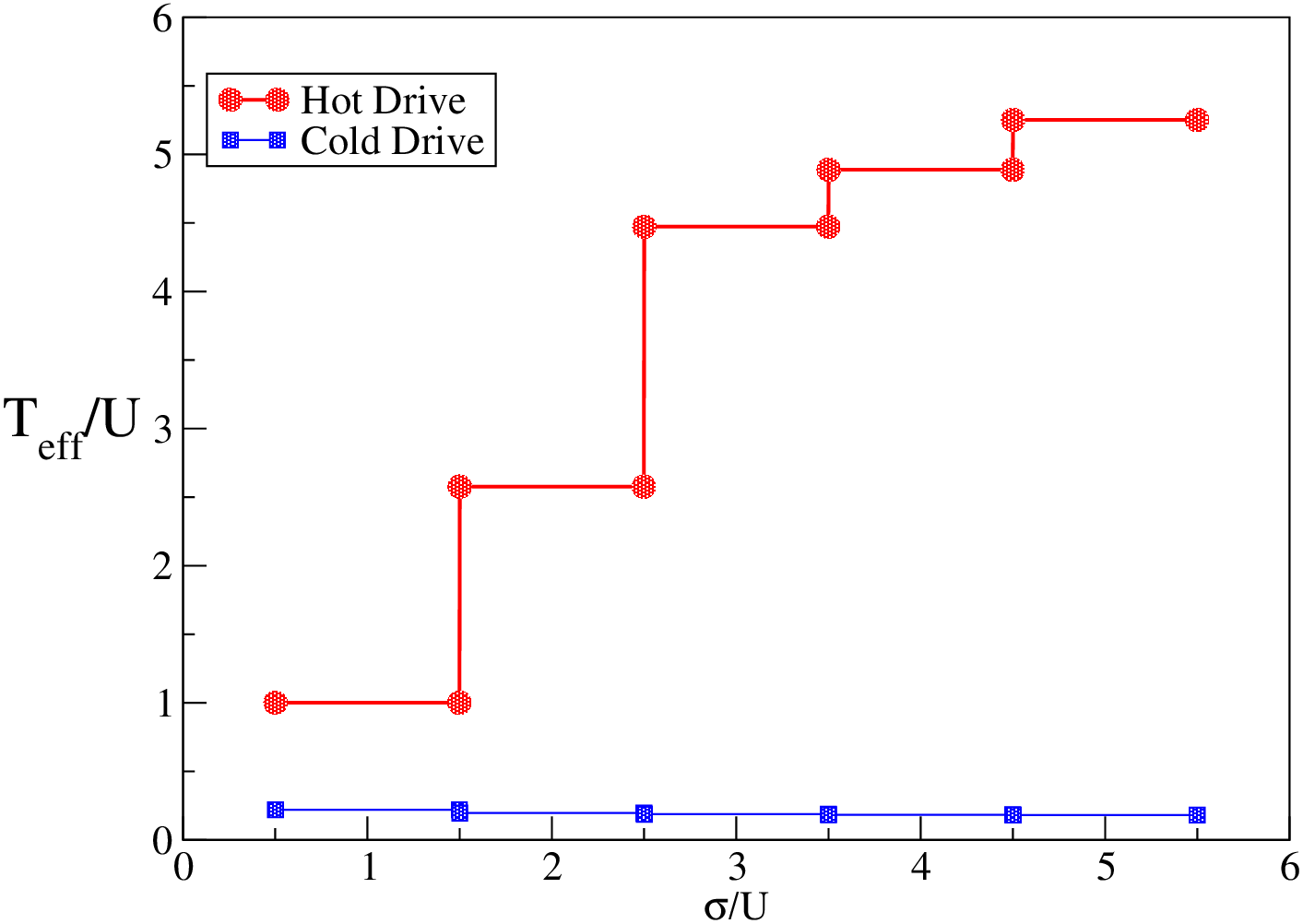}
\caption{Effective temperature $T_{eff}$ as a function of the drive bandwidth $\sigma$ for  the two driving protocols considered. We notice that while in the cold drive case $T_{eff}$ depends only weakly on $\sigma$ and stay small, for the hot drive case it substantially increases as $\sigma$ (and therefore the number of photons) increases.
 Parameters: loss rate $\kappa=0.0128 U$, resonator frequency $\delta\omega_0=0.0$, drive bandwidth (left panel) $\sigma= 0.625U$, drive amplitude (right panel) $f=0.0625 U$. }
\label{fig:fig7}
\end{figure}
\begin{figure}[h!]
\includegraphics[width=8.cm]{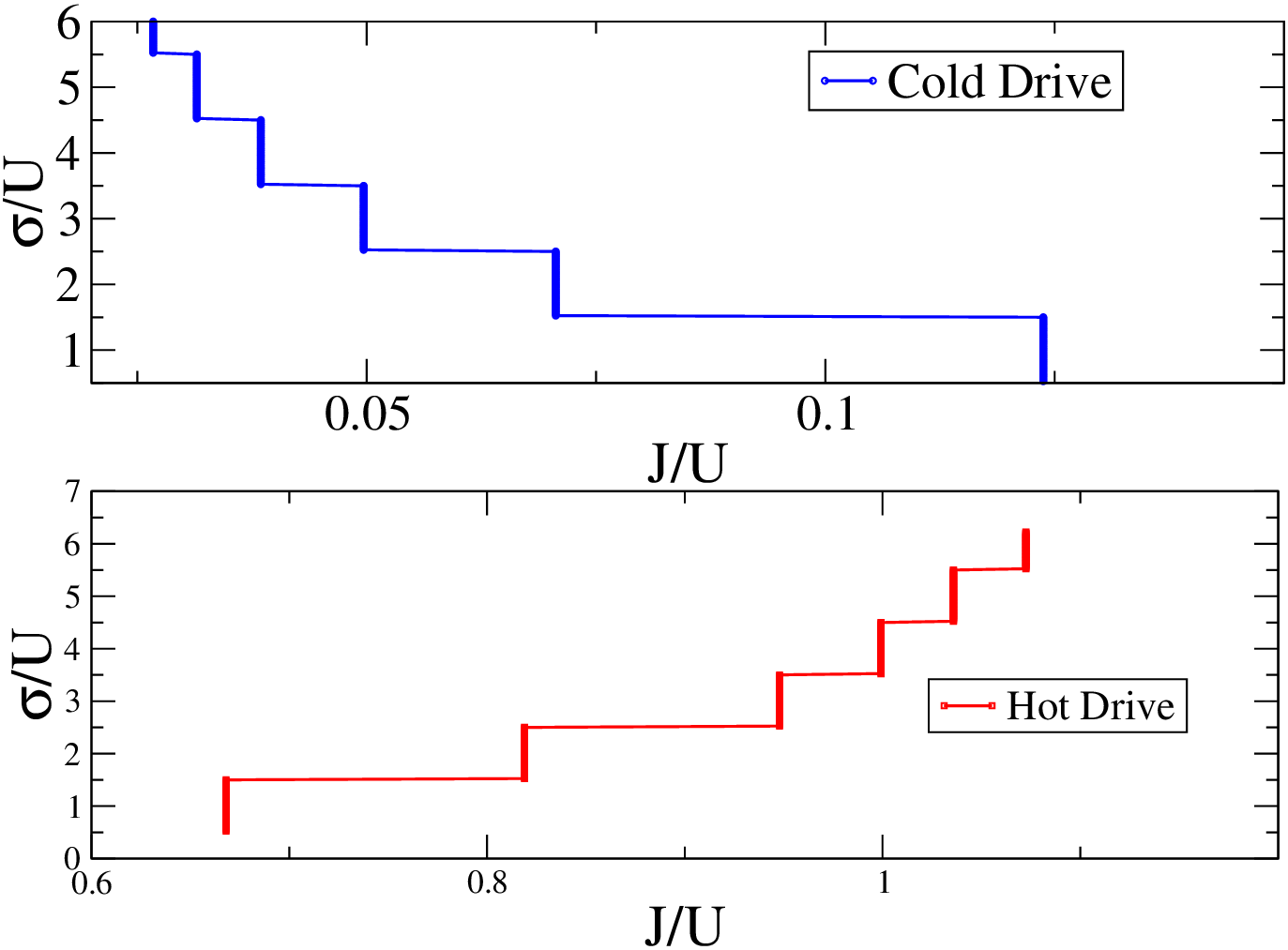}
\caption{Stationary State Phase diagram as a function of hopping strenght $J$and drive bandwidth $\sigma$, for the two driving protocols considered namely cold drive (top panel) and hot drive (bottom panel). Parameters: loss rate $\kappa=0.0128 U$, resonator frequency $\delta\omega_0=0.0$, drive   amplitude (left panel) $f=0.0625 U$.}
\label{fig:fig8}
\end{figure}

\subsection{Local Bosonic Occupation and Effective Temperature}

As we discussed in section \ref{sec:modelAndDriving}, figure~\ref{fig:fig1}, the boson number in the single-site problem as a function of pump bandwidth $\sigma$, shows a starcaise structure characteristic of blockade physics~\cite{Houck_prl11}, with a value of $\Delta\sigma\sim U$ required to add extra bosons in the system.
In this respect,  an important difference between the two driving protocols already appear, namely the cold drive is able to fix the occupancy to almost integer filling~\cite{LebreuillyEtAlPRA17}, while the hot drive to half-integer filling, reflecting the fact that the stationary density matrix is almost pure in the cold drive case, while it has a box-shaped distribution of populations in the hot case.
The dependence of $T_{eff}$, introduced in section~\ref{sec:singleSite}, from the drive bandwidth $\sigma$ is plotted in figure~\ref{fig:fig7} and reveals a rather substantial difference in the two driving protocols for what concerns the effective heating properties of the system. Indeed upon increasing the drive bandwidth the effective temperature increases in the hot drive case while decreseas (slightly) in the cold case one. For the hot drive, this can be understood because the density matrix would reach an infinite temperature one by sending the bandwith of the drive to infinity. This offers an alternative perspective on the recent proposed scheme to engineer effective ground state phases of interacting photons through the use of non-markovian reservoirs~\cite{LEBREUILLY2016836,LebreuillyEtAlPRA17}.

\subsection{Phase Diagram}

Finally we conclude presenting, in figure~\ref{fig:fig8}, the stationary phase diagram of the finite-frequency phase transition for the two driving protocols we have discussed so far. In the main text we considered a specific value of the drive bandwidth $\sigma$  ($\sigma =1.5 U$) while now we present the phase boundary in the $\sigma,J$ plane.  As we discussed in the main text $\sigma$ controls both the local density and the effective chemical potential $\Omega_*$, in step like fashion that resembles the equilibrium ground-state physics of the problem. It is therefore a natural choice for a tuning parameter in the phase diagram. For both driving protocols we generically find a similar behavior, namely a small hopping phase $J<J_c(\sigma)$ which has a stable stationary state fully incoherent and a large hopping regime $J>J_c(\sigma)$ where the stationary state becomes unstable toward an oscillating regime and the system develops a $U(1)$ order parameter at finite frequency. It is nevertheless quite interesting to discuss the different shapes of the phase boundary, which instead rather strongly depend on the protocol. We notice that for the cold drive case(top panel) the boundary resembles the ground state one, with a lobe-like structure for different values of the local density and a critical hopping $J_c$ which decreases as the local filling increases. Viceversa, in the hot drive case (bottom panel) we find a rather opposite effect, namely the critical hopping increases with $\sigma$ and the region of normal phase stability expands. We can understand this effect from the discussion on the occupation of the bosonic mode and the effective temperature: indeed in the hot drive case, increasing the bandwidth $\sigma$ has the effect of both changing the local occupation (see figure~\ref{fig:fig1}) and of increasing the effective temperature (see figure~\ref{fig:tempVsSigma}), with the result of shrinking the broken symmetry region due to effectively increased thermal fluctuations.

\end{document}